# Forty years of the Ellis–Baldwin test

Nathan Secrest[1†], Sebastian von Hausegger[2], Mohamed Rameez[3], Roya Mohayaee[4] & Subir Sarkar[2]

[1] Celestial Reference Frame Department, U.S. Naval Observatory, Washington, DC, USA.
[2] Department of Physics, University of Oxford, Oxford, UK.
[3] Department of High Energy Physics, Tata Institute of Fundamental Research, Mumbai, India.
[4] Sorbonne Université, CNRS, Institut d'Astrophysique de Paris, Paris, France.

[†]email: nathan.j.secrest.civ@us.navy.mil

**Modern cosmology is built on the assumption that the Universe is homogeneous and isotropic on large scales — but this is challenged by results of the Ellis–Baldwin test that show an unexplained anomaly in the distribution of distant galaxies and quasars.**

In the standard model of cosmology, known as 'Lambda Cold Dark Matter' (ΛCDM), it is assumed that spacetime is described by the homogeneous and isotropic Robertson–Walker metric. This assumption, called the cosmological principle, allows a simple solution to the Einstein field equations — the Friedmann–Lemaître equations — that describes the evolution of the Universe and includes a so-called cosmological constant Λ. In the present epoch of an apparently flat Universe, the energy density of matter is ~30% and the energy density associated with Λ is ~70%: the Universe is dominated by the cosmological constant, also called dark energy.

Onto this framework are imposed scalar density perturbations from which structure forms at later epochs through gravitational instability. These perturbations are imprinted as small-scale temperature anisotropies (of order $10^{-5}$) on the cosmic microwave background (CMB) at the surface of last scattering around redshift $z \sim 1{,}100$. The structure of the Universe at later epochs can then be predicted using linear perturbation theory and the cosmological parameters estimated by fitting the power spectrum of CMB anisotropies and large-scale structure. In this model framework, significant deviations from isotropy and homogeneity can therefore arise only after structure begins to form.

Testing for such deviations, especially with respect to isotropy, was the motivation for a 1984 paper[1] by the relativist George Ellis and radio astronomer John Baldwin. In 1967 it had been noted[2] that deviations from the smooth Hubble flow on small scales due to structure formation should give rise to a dipole anisotropy in the CMB (of the form shown in Fig. 1) with amplitude $\mathcal{D} = v/c \sim 10^{-3}$, where $v$ is the observer's velocity with respect to the cosmic rest frame (in which the Universe looks isotropic) and $c$ is the speed of light. The predicted CMB dipole had subsequently been detected; the most recent measurement by the Planck satellite[3] indicates that the velocity of the Solar System is $(369.82 \pm 0.11)$ km s$^{-1}$, in the direction $l = 264.021 \pm 0.011$, $b = 48.243 \pm 0.005$ in Galactic coordinates.

Ellis and Baldwin[1] estimated that this motion might also be detectable in future catalogues of radio sources. These sources, expected to be found near $z \sim 1$ and numbering in the hundreds of thousands, would exhibit a dipole anisotropy with a relative amplitude $\mathcal{D} = [2 + x(1 + \alpha)] \, v/c$, where $x$ is the index of the integral-source-counts power law and $\alpha$ is the source spectral index. The Ellis–Baldwin test is elegant in its simplicity: it is valid for any set of cosmologically distant, discrete sources for which the experimenter has measured fluxes and spectral indices, and it is model-independent, relying only on special relativity. In particular, the Ellis–Baldwin test is robust against any evolution of the sources with redshift if measurements of $x$ and $\alpha$ are made at the threshold of a flux-limited survey[4]. Hence it is a powerful consistency test of the Friedmann–Lemaître–Robertson–Walker cosmology.

The main difficulty of the Ellis–Baldwin test, however, is that it requires very large and clean catalogues of extragalactic sources to detect the predicted 370 km s$^{-1}$ kinematic dipole above random fluctuations induced by shot noise. Ellis and Baldwin estimated that detecting the expected relative amplitude for radio sources of 0.0046 at the 3σ significance level would require ~ 200,000 sources — which did become available 14 years later with the publication of the National Radio Astronomy Observatory Very Large Array Sky Survey (NVSS). In 2002, using NVSS data, Chris Blake and Jasper Wall announced[5] that they had detected the expected dipole, an apparent confirmation that the velocity dipole of extragalactic matter is consistent with that of the CMB.

**A too-large matter dipole**

Unfortunately, however, Ellis and Baldwin's signal-to-noise calculations were optimistic. Indeed, it was shown[6] that a radio survey such as the NVSS would in fact require not 200,000 but about two million sources to detect the expected dipole at the 3σ level, making a detection using the sample constructed by Blake and Wall impossible. Nonetheless, Blake and Wall had detected a dipole anisotropy in radio source counts, achieving a signal-to-noise ratio of ~3.7 — made possible only because the corresponding velocity was (900 ± 250) km s$^{-1}$, a 2.1σ tension with the predicted value. Later investigations also found a too-large dipole in NVSS data using a variety of flux cuts, masking strategies, and estimators. These studies consistently concluded that there is a detectable dipole in catalogues of radio galaxies, aligned close to the dipole of the CMB, but several times larger than the kinematic expectation.

A spuriously large dipole can most readily be attributed to the contribution of clustering from local sources. Indeed, although the extragalactic radio population is, a priori, expected to be at moderate redshifts ($z >$ 0.1), the redshift distribution of these sources is empirically poorly constrained. There are good reasons to be sceptical, however, that there is an underestimated population of local radio galaxies contributing a larger-than-expected clustering signal to kinematic dipole estimates. For example, in constructing their sample Blake and Wall[5] first removed local galaxies that comprise the 'clustering dipole', estimating that such sources are only important at redshifts less than 0.03. Colin et al.[7] constructed a nearly full-sky radio galaxy sample using the NVSS and the Sydney University Molonglo Sky Survey, testing the effect of removing as many potential local contaminants as possible (such as cutting out the supergalactic plane, removing large superclusters, and excluding matches to the 2MASS Redshift Survey). These tests showed that the effect of these sources on the radio galaxy dipole cannot account for the anomalously large values found.

Perhaps the simplest objection to attributing the radio galaxy dipole to local clustering is the observation that such nearby galaxies should have been detected in wide-area surveys such as the Sloan Digital Sky

Survey. It is relevant to note however that there are unexpectedly fast and deep coherent peculiar motions, or bulk flows, in the local Universe[8]. If these are due to inhomogeneities, the corresponding clustering dipole may be higher than expected in ΛCDM. However this cannot be the cause of the too-large dipole seen in quasars (discussed below) which are more certainly at higher redshift.

**Independent confirmation**

The dipole anomaly remained at relatively low statistical significance (~2–3σ) for nearly two decades, mainly owing to there being limited radio data suitable for the Ellis–Baldwin test, so was largely ignored. Even with a higher formal significance, however, systematic errors are an ever-present concern. Radio survey data is prone to declination-dependent systematics, loss of completeness around bright sources, and flux calibration systematics. Although Ellis and Baldwin envisaged radio galaxies as the cosmological test particles needed to probe for large-scale departures from isotropy and homogeneity, what was really needed was independent confirmation of the kinematic dipole tension that radio studies hinted at.

This became possible with a new catalogue produced using mid-infrared data from the Wide-field Infrared Survey Explorer (WISE), launched in 2009, which performed a survey of the full sky at wavelengths of 3.4, 4.6, 12, and 22 microns. Mid-infrared photometry is especially efficient at separating different classes of astrophysical objects, especially stars from extragalactic sources, and WISE data have been used successfully to produce catalogues of millions of active galactic nuclei and quasars. Using WISE data, we constructed a sample of 1.36 million quasars with median redshift of $z \sim 1.2$ covering ~50% of the sky, enabling the most precise measurement of the kinematic dipole of moderate redshift matter to date[9]. We found that, like the radio-galaxy dipole, the quasar dipole is aligned closely with that of the CMB, but the amplitude, corresponding to ~800 km s$^{-1}$, is more than twice as large as the kinematic expectation. Moreover, simulations show that if the Solar system is indeed moving at 370 km s$^{-1}$ with respect to an isotropic distribution of quasars, this result would be expected by chance only once in two million times.

This result, corresponding to a 4.9σ disagreement with the standard kinematic expectation, was the most significant confirmation of the dipole anomaly ever published at the time. However just as important is the fact that this result was achieved using systematically independent data. Data from WISE share no instrumental, observational, or astrophysical commonalities with ground-based radio data. The survey strategies are different, as are the flux calibrations, wavelengths, astrophysical foregrounds, and even source type, with fewer than 2% of WISE-selected quasars having counterparts in the NVSS above flux densities of 10 mJy. We leveraged this near-total independence of quasars and radio galaxies in a follow-up joint analysis of these objects, resulting in a 5.1σ disagreement with the kinematic expectation[10].

It is this systematically independent confirmation, more than the formal ~5σ formal significance, that convincingly suggests there is something strange about the Universe at moderate redshift. Simply stated, it is difficult to believe that the independent systematics of ground-based radio data and space-based infrared data could conspire to give the same answer. Additionally, more recent radio data, such as from the Rapid ASKAP Continuum Survey (ASKAP is the Australian Square Kilometer Array Pathfinder), have now enabled an independent 4.8σ confirmation[11] of the anomaly, putting to rest any lingering concerns about the formal significance based only on radio data. It seems that the dipole apparent in radio galaxies and quasars is a reality, one that requires either that the rest frame of large-scale structure be different from that of the CMB or that the distribution of matter on large scales is less homogeneous than predicted by ΛCDM. Either way, the conclusions drawn about the Universe in the standard cosmological model may need re-examination.

**More data to challenge the model**

One might consider the magnitude of the dipole anomaly (~1%) and wonder just how important it really is for the foundations of modern cosmology. Perhaps the cosmological principle is a good enough approximation to reality. Indeed, one might argue that the usual assumption of Gaussian, near scale-invariant adiabatic density perturbations in the early Universe is not accurate enough. This, however, should be manifest in the CMB power spectrum — but as discussed above it is the CMB power spectrum that sets the cosmological parameters that in turn determine the scale of homogeneity, and non-Gaussianities have not been yet detected. It would thus seem that, under the standard cosmology, one can either accept the predictions of the standard cosmological model based on the CMB power spectrum or one can accept the measured moderate-redshift dipole, but not both.

One way or the other, isotropy and homogeneity are simplifying assumptions — made at a time when there were almost no data — that enabled the construction of a cosmological model built on general relativity. This model has undoubtedly been useful, but all models are wrong in an absolute sense, as the statistician George E. P. Box emphasized. Thus ΛCDM is not unassailable in the face of new data: it should be expected to give way eventually to a deeper and more explanatory model, especially given that there is no physical understanding of Λ.

The prospects for such a paradigm change in the near future are in fact rather good. In the next few years, several new ground- and space-based observatories will survey the sky in unprecedented detail, providing high-precision data on millions of galaxies and quasars. Surveys such as the Vera C. Rubin Observatory's Legacy Survey of Space and Time and the Dark Energy Spectroscopic Instrument survey, and those undertaken by ESA's Euclid, NASA's SPHEREx and the Square Kilometer Array, will dramatically increase our understanding of the moderate-redshift Universe, setting critical new constraints on cosmological models. The nature of the kinematic dipole anomaly, which challenges the foundations of modern cosmology, might soon be revealed.

**Fig. 1 | The dipole in the cosmic microwave background compared with the dipole in the sky map of quasars, which is about twice as large as expected.** The image of the microwave sky at 70 GHz (left) is by the Planck satellite (https://beyondplanck.science), while the smoothed map of quasars at 3.4 μm (right) was taken by the WISE satellite (Secrest et al. 2022).

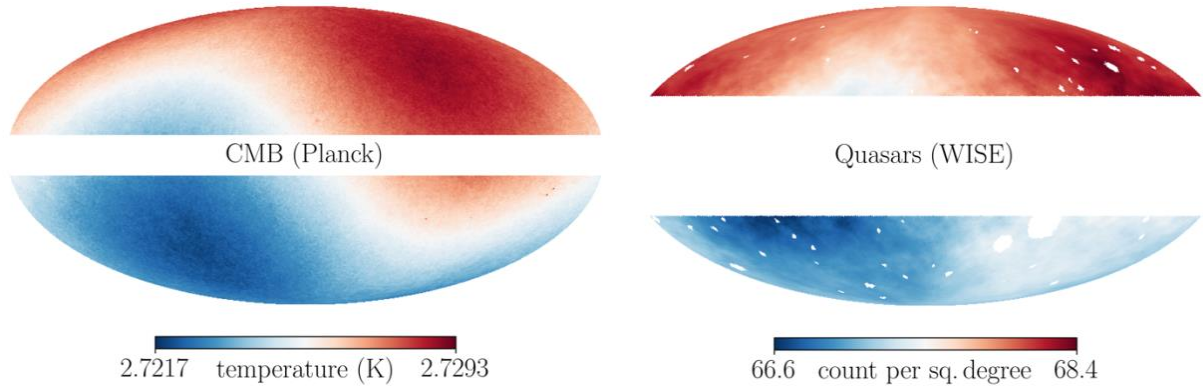